\documentclass[journal]{IEEEtran}

\usepackage{ifpdf}
\usepackage{cite}
\usepackage{amsmath}
\usepackage{algorithm}
\usepackage{threeparttable}
\usepackage[noend]{algpseudocode}
\usepackage{algorithmicx}
\usepackage{array}
\usepackage{epsfig}
\usepackage{graphicx}
\usepackage{subfigure}
\usepackage{epstopdf}
\usepackage{amssymb}
\usepackage{array}
\usepackage{color}
\usepackage{booktabs}
\usepackage{multirow}
\usepackage{verbatim}

\usepackage[]{pifont}

\begin{document}

\title{Probability-based Distance Estimation Model for 3D DV-Hop Localization in WSNs}


\author{Penghong~Wang, Hao Wang, Wenrui Li,
	Xiaopeng~Fan,~\IEEEmembership{Senior Member,~IEEE,} and Debin~Zhao,~\IEEEmembership{Member,~IEEE}
        
        
\thanks{This work was supported in part by the National Key Research and Development Program of China (2021YFF0900500), and the National Science Foundation of China (NSFC) under grant U22B2035.}

\thanks{Penghong Wang and Wenrui Li are with the Faculty of Computing, Harbin Institute of Technology, Harbin 150001, China. e-mail: (phwang@hit.edu.cn; 21B903007@stu.hit.edu.cn). }

\thanks{Hao Wang is with the Faculty of Computing, Harbin Institute of Technology, Harbin 150001, China, and also with the College of Engineering, City University of Hong Kong, Kowloon 999077, Hong Kong. e-mail: (ho.wong@cityu.edu.hk).}

\thanks{Xiaopeng Fan and Debin Zhao are with the Faculty of Computing, Harbin Institute of Technology, Harbin 150001, China, and also with the PengCheng Lab, Shenzhen 518055, China. e-mail: (fxp@hit.edu.cn; dbzhao@hit.edu.cn). }

\thanks{Corresponding author: Xiaopeng Fan.} 
}

\markboth{}%
{Shell \MakeLowercase{\textit{et al.}}: Bare Demo of IEEEtran.cls for IEEE Journals}
%

\maketitle

\begin{abstract}

Localization is one of the pivotal issues in wireless sensor network applications. In 3D localization studies, most algorithms focus on enhancing location prediction process, lacking theoretical derivation of the detection distance of an anchor node at the varying hops, which engenders a localization performance bottleneck. To address this issue, we propose a probability-based average distance estimation (PADE) model that utilizes the probability distribution of node distances detected by an anchor node. The aim is to mathematically derive the average distances of nodes detected by an anchor node at different hops. First, we develop a probability-based maximum distance estimation (PMDE) model to calculate the upper bound of the distance detected by an anchor node. Then, we present the PADE model  relies on the upper bound obtained of the distance by the PMDE model. Finally, the obtained average distance is used to construct a distance loss function, and it is embedded with the traditional distance loss function into a multi-objective genetic algorithm to predict the locations of unknown nodes. The experimental results demonstrate the proposed method achieves the state-of-the-art performance in both random and multimodal distributed sensor networks. The average localization accuracy is improved by 3.49\%-12.66\% and 3.99\%-22.34\%, respectively.

\end{abstract}


\IEEEpeerreviewmaketitle

\section{Introduction}

\IEEEPARstart{W}{ireless} sensor networks (WSNs) have shown appealing powers in industry \cite{tiwari2007energy, Hodge2015, Kaur2022}, environmental monitoring \cite{Air02}, healthcare \cite{Reyouchi2022, Zhu2022}, and supply chain management \cite{wang2015wireless}. WSNs are composed mainly of a large number of tiny sensors in an ad-hoc manner. These sensors are deployed in a physical region arbitrarily to monitor and manage environmental conditions in a specific area or to perform certain tasks of collecting information. These tasks typically require obtaining sensor location information, without accurate location information, sharing the data collected by sensors will be inefficient or even infeasible. Global Positioning System (GPS) can provide sensors with the ability to be aware of location in real-time, but installing GPS on all sensors in a large-scale WSN will significantly increase the cost and energy consumption of the system. In contrast, the range-free localization scheme is well recognized as a promising and low-energy consumption approach for wireless sensor localization.

\begin{figure}[t]
	
	\centerline{\includegraphics[scale=0.7]{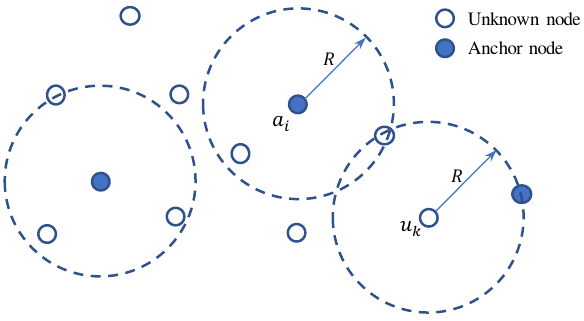}}
	\caption{An example of deploying sensor nodes in a wireless sensor network.  $ a_i $: anchor node; $ u_k $: unknown nodes.}
	\label{BG}
\end{figure}

The range-free localization schemes\cite{niculescu2003dv09}\cite{Wang2009, Xiao2010, Bang2012A, Slim2016, Cui2018, WangHP2019, liuGui2019, caiSC2020, Chen2020, Liu2020, WangKS2022, Chen2022, Wang2023}\cite{shahzad2017, CuiJ2017, GuiTVT2020, WangCCPE2020, kanwar2020, OUYANG2021, jinWCL2022, WangSJ2022, LIUN2022, JIA2022, Cao2023} \cite{TanTSN2013, CuiTIM2016, Fan2017, Liu2019, caiS2019, kanwar2021, Sah2022, Abuaddous2023, Cheng2023, WangSC2023} is one of the prevailing methods in wireless distributed multi-agent localization. Its distinguishing feature lies in its ability to locate without the need for precise measurements of distances or angles between nodes.  It can predict the locations of unknown nodes only by relying on the location coordinates of a few nodes and the number of connections between nodes. Fig. \ref{BG} shows an example of deploying sensor nodes in a wireless sensor network, including anchor nodes (with known location information) and unknown nodes (with unknown location information). Based on the location of node $ a_i $ and the hop counts between nodes, the distance from node $ u_k $ to node $ a_i $ can be estimated, then the location of node $ u_k $ can be predicted. Especially, the distance vector hop (DV-Hop) localization algorithm is one of the most representative range-free localization schemes.

DV-Hop localization algorithm was proposed by Niculescu \textit{et al.}\cite{niculescu2003dv09} in 2003, and its positioning process contains three parts. \textbf{1)} Obtain the minimum connections between nodes based on message passing; \textbf{2)} Calculate the Euclidean distance between an unknown node and an anchor node based on the locations of anchor nodes and the minimum connections; \textbf{3)} Predict the geographic location of each unknown node based on the information obtained from the above process. The DV-Hop algorithm provides a simple and low-power localization scheme, but its accompanying issue is a large localization error. However, the advantage of its low-cost positioning still attracts scholars to continuously ameliorate the algorithm.

Based on the DV-Hop algorithm, various improved solutions are proposed to solve wireless sensor localization tasks in two-dimensional (2D) or three-dimensional (3D) spatial scenes. Specifically, most improvement strategies mainly focus on localization issues in 2D planar scenes. For example, some works \cite{Wang2009, Xiao2010, Bang2012A, Slim2016, Cui2018, WangHP2019, liuGui2019, caiSC2020, Chen2020, Liu2020, WangKS2022, Chen2022, Wang2023} focus on providing more accurate solutions for estimating the distance between an unknown node and an anchor node. Especially, \cite{ Wang2023} provides a distance derivation model for 2D localization scenarios. Others \cite{shahzad2017, CuiJ2017, GuiTVT2020, WangCCPE2020, kanwar2020, OUYANG2021, jinWCL2022, WangSJ2022, LIUN2022, JIA2022, Cao2023} focus on exploring various location prediction schemes to provide more accurate accuracy for unknown nodes. In recent years, with the deployment and application of wireless sensors in complex terrain such as mountains, underwater, and tunnels, the issue of wireless sensor localization in 3D spatial scenes has gradually attracted attention. Some methods\cite{TanTSN2013, CuiTIM2016, Fan2017, Liu2019, caiS2019, kanwar2021, Sah2022, Abuaddous2023, Cheng2023, WangSC2023} are proposed to promote positioning performance in 3D spatial scenes, such as connectivity-based and anchor-free \cite{TanTSN2013}, distributed algorithms \cite{Fan2017} and  multi-objective optimization methods\cite{caiS2019, kanwar2021, WangSC2023}. These works primarily focus on enhancing localization algorithms, but lack a theoretical derivation of the detection distance for an anchor node.

In this paper, motivated by \cite{ Wang2023} for 2D DV-Hop localization, we propose a probability-based maximum distance estimation model and an average distance estimation model for 3D DV-Hop localization. The aim is to explore the correlation among distance estimation, hop counts, and the number of detecting nodes in 3D environments, enhance the overall localization performance. The main contributions are summarized as follows:

\begin{enumerate}
	\item [\textbf{1)}] We propose a probability-based maximum distance estimation (PMDE) model to calculate the upper bound of the distance detected by an anchor node. The PMDE model is constructed based on the probability distribution of the distance between each unknown node and an anchor node under different hops.

	\item [\textbf{2)}] Based on the upper bound obtained of the distance by the PMDE model, we propose the probability-based average distance estimation (PADE) model to calculate the average distance detected by an anchor node at different hops.

	\item [\textbf{3)}] We establish a distance loss function based on the obtained expected distance and integrate it with the existing distance loss function into a multi-objective genetic algorithm to accurately predict the location of each unknown node. The numerical results demonstrate significant gains of our approach compared to the state-of-the-art techniques.

\end{enumerate}

The rest of the paper is organized as follows: Section II reviews the improved DV-Hop algorithm in 3D localization scenarios. Section III proposes a probability-based maximum distance estimation model. Section IV proposes a probability-based average distance estimation model. Section V constructs two distance loss functions and embeds them into the multi-objective optimization algorithm. Section VI discusses the simulation results. Finally, the conclusion of this paper is summarized.

\section{Related Work}

\subsection{3D DV-Hop }

The 3D DV-Hop algorithm is conceptually akin to its 2D counterpart. Firstly, the average hop distance for each anchor node is determined based on the minimum hops obtained through the flooding process of the anchor nodes, and it is calculated by
\begin{equation}
	A\_Dis_{i} = \frac{{\sum\limits_j {\sqrt {{{({x_i} - {x_j})}^2} + {{({y_i} - {y_j})}^2} + {{({z_i} - {z_j})}^2}} } }}{{\sum\limits_j {Ho{p_{i,j}}} }},
	\label{eq0}
\end{equation}
where $(x_{i}, y_{i}, z_{i})$ and $(x_{j}, y_{j}, z_{j})$ indicate the coordinates of $a_{i}$ and $a_{j}$, respectively, and $Hop_{i,j}$ indicates the minimum connections between $a_{j}$ and $a_{i}$. The estimated distance between $u_{k}$ and $a_{i}$ is calculated by 
\begin{equation}
{Dis}_{i, k} = A\_Dis_{i} \cdot {Hop}_{i, k},
\label{eq01}
\end{equation} 
where $Hop_{i,k}$ stands for the minimum connections between $u_{k}$ and $a_{i}$. Finally, the location of each $u_{k}$ can be predicted by the least mean square estimator.

\subsection{3D DV-Hop-based Improved Sschemes}

In 2D positioning scenarios, the distance estimation strategies \cite{Cui2018, Wang2023, caiSC2020} and the position optimization strategies\cite{CuiJ2017, GuiTVT2020, OUYANG2021} have significantly advanced the progress of localization technologies. However, in real-world applications, when sensors are deployed in areas with varying terrain and elevation, the positioning system must be able to accurately determine the vertical and horizontal locations of nodes. In this case, it is necessary to develop 3D positioning methods to effectively overcome the limitations of 2D positioning methods in the vertical dimension. \cite{TanTSN2013} proposes a connectivity-based and anchor-free localization scheme, aiming to address the challenge of node localization in large-scale 2D/3D sensor networks with concave areas. This method can achieve smooth expansion from 2D to 3D. \cite{Fan2017} proposes a distributed algorithm to achieve accurate 3D sensor localization. Specifically, the proposed scheme utilizes connected information to implement an approximate convex partitioning technique that divides the entire network into multiple subnetworks. Subsequently, the multidimensional scaling-based algorithm is employed to precisely locate nodes within each subnetwork. \cite{caiS2019} constructs a distance loss function based on the average distance detected per hop of all anchor nodes and employs a multi-objective localization scheme to optimize the predicted node positions. This work is the first to apply multi-objective optimization algorithms to 3D positioning problem. \cite{kanwar2021} develops a localization scheme using multi-objective particle swarm optimization. It further enhances the application of multi-objective optimization algorithms in 3D sensor localization problems. \cite{Cheng2023} presents a maximum similarity path method for estimating the distance between an unknown node and an anchor node and utilizes the cosine theorem to correct the estimated distance. Additionally, an improved water flow optimizer is employed to predict the locations of the unknown nodes.

These works primarily focus on enhancing localization algorithms, but lack a theoretical derivation of the detection distance for an anchor node. In this work, we propose a probability-based maximum distance estimation model and an average distance estimation model for 3D DV-Hop localization, which extends the distance estimation model from 2D localization scenes in \cite{ Wang2023} to 3D localization scenes.

\section{Probability-based Maximum Distance Estimation}

\begin{figure}[htbp]
	
	\centerline{\includegraphics[scale=0.6]{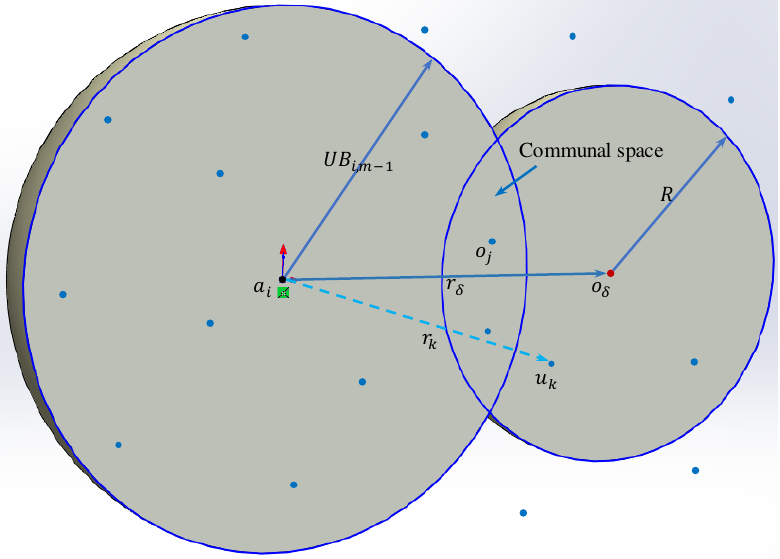}}
	\caption{The example of multi hop information transmission for sensors. $ {a}_{i} $: an anchor node, $ {u}_{k} $: an unknown node, $ {o}_{j}$ and $ {o}_{\delta}$: an ordinary (anchor or unknown) node, $ {r}_{\delta}$: the distance between $ {o}_{\delta}$ and $ {a}_{i} $, $ UB_{i,m-1} $: the expected distance of the outermost node detected by node $ {a}_{i} $ when $ hop=m-1 $.}
	\label{hopm}
\end{figure}

Our goal is to estimate the average distance detected by an anchor node at different hops for locating each unknown node position. However, the computation of the average distance requires obtaining the upper bound of the distance detected by an anchor node. To adress this issue, we propose a probability-based maximum distance estimation model for 3D sensor localization scenarios in this section. Its aim is to provide accurate distance estimation for the outermost detection node that is detected by anchor nodes at different hop counts. Fig. \ref{hopm} provides a visualization example that represents the information transmission process of sensor nodes.

Assuming that the deployment of sensor nodes in 3D space follows a uniform distribution, as shown in Fig. \ref{hopm}. $ {a}_{i} $ is an anchor node, $ {o}_{j}$ and $ {o}_{\delta}$ are ordinary nodes (anchor or unknown nodes), $ {r}_{\delta}$ indicates the distance between $ {o}_{\delta}$ and $ {a}_{i} $, and $ UB_{i,m-1} $ is the distance of the outermost node detected by $ {a}_{i} $ when $ hop=m-1 $. In Fig. \ref{hopm}, let $ {o}_{\delta}$ be the outermost node detected by $ {a}_{i} $ when $ hop=m $. In this case, two conditions need to be satisfied: \textbf{1)} $ {o}_{\delta}$ is the outermost node within the detection range of $ {a}_{i}$ when $ hop=m $; \textbf{2)} the communal space contains at least one node relay node (denoted as $ {o}_{j}$). Based on the above analysis, let $ P_{i, m-1}(o_j) $ represent the probability that there exists at least one $ {o}_{j}$ in the communal space, let $ P_{i,m}(o_\delta) $ represent the probability that $ {o}_{\delta}$ is the outermost node within the detection range of $ {a}_{i}$ when $ hop=m $. The probability density ($ P_{i, m}^{true}(o_\delta) $) of the outermost node that $ {a}_{i} $ can detect when $ hop=m $ is calculated by
\begin{equation}
P_{i, m}^{true}(o_\delta)=P_{i, m-1}(o_j)P_{i,m}(o_\delta).
\label{eq2}
\end{equation}

\begin{figure}[htbp]
	
	\centerline{\includegraphics[scale=0.63]{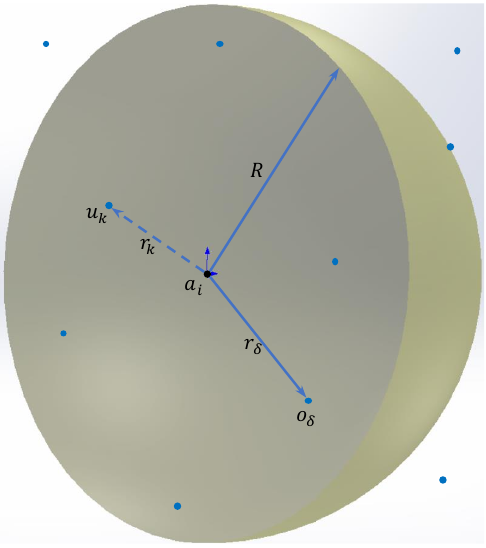}}
	\caption{The example of node distribution detected by $ a_i $ when $ m=1 $.}
	\label{hop1}
\end{figure}

Next, we introduce the calculation methods of $ P_{i, m-1}(o_j) $ and $ P_{i, m}(o_\delta) $ involved in (\ref{eq2}). Significantly, when $ m=1 $, the distribution of nodes is shown in Fig. \ref{hop1}. Let $ o_\delta $ represent the outermost node detected by $ a_i $ when $ m=1 $, the  probability $ P_{i,m}(o_\delta) $ is calculated by
\begin{equation}
\begin{aligned}
P_{i, 1}(o_\delta)
&=\prod_{k=1}^{n_{i,1}}{P_{i,1}(r_\delta \geq r_k)}
\\
&=\prod_{k=1}^{n_{i,1}}{ \frac{\int_{0}^{r_\delta}{(R^2-r^2)dr}}{\int_{0}^{R}{(R^2-r^2)dr}} }
\\
&=\left(\frac{{r_\delta}^3}{R^3}\right)^{n_{i,1}}
\end{aligned}, m=1,
\label{eq3}
\end{equation}
where $ r_\delta $ represents the distance between $ o_\delta $ and $ a_i $, and $ r_\delta=max(r_1,r_2,...,r_k,...,r_{n_1}) $, $ n_{i,1} $ represents the number of nodes detected by $ a_i $ when $ hop=1 $. In this scenario, the nodes detected by $ a_i $ are independent of any relay nodes, i.e., $ P_{i, m-1}(o_j)=1$, and (\ref{eq2}) can be simplified as $ P_{i,1}^{true}(o_\delta)=P_{i,1}(o_\delta) $.

\begin{figure*}[t]
	
	\center
	\subfigure[]{\includegraphics[scale=0.6]{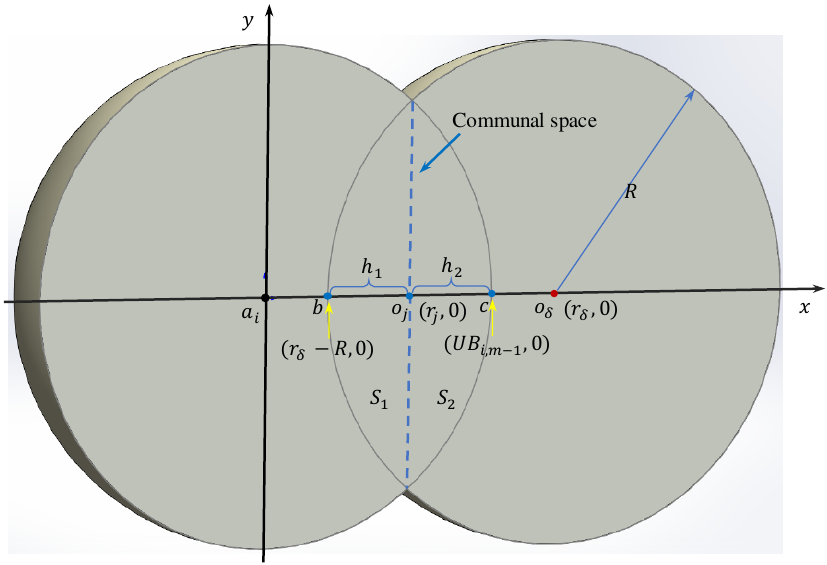}}~~~
	\subfigure[]{\includegraphics[scale=0.57]{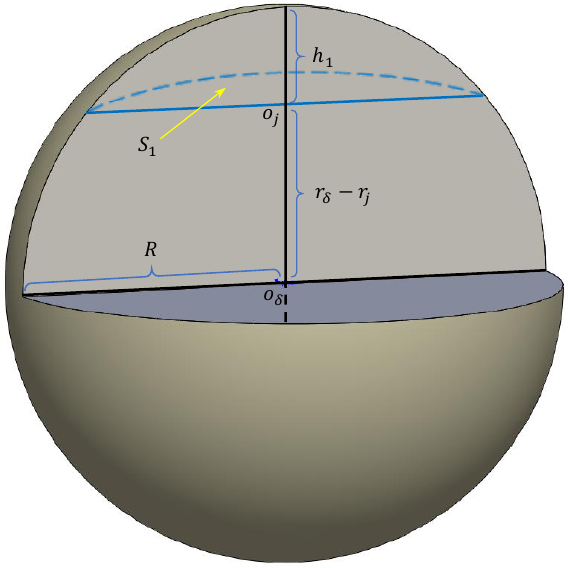}}
	
	\caption{An example of outermost detection node analysis. $ {a}_{i} $: an anchor node, $ o_\delta $: an ordinary node, $ S_1 $ and $ S_2 $: a part of the communal space, $ R $: communication radius,  $ UB_{i,m-1} $: the expected distance of the outermost node detected by node $ {a}_{i} $ when $ hop=m-1 $.}
	\label{joint}
\end{figure*}

When $ m>1 $, we model the distribution of nodes and the multi-hop transmission characteristics of information in a Cartesian coordinate system, as depicted in Fig. \ref{joint}. In Fig. \ref{joint} (a), $ a_i $ denotes the center of the Cartesian coordinate system, $ o_\delta $ denotes the outermost node detected by $ a_i $ when $ hop=m $, its coordinates are $ (r_\delta,0) $. $ o_j $ is Located at the center of the communal space, with coordinates $ (r_j,0) $. The coordinates of nodes $ b $ and $ c $ are $ (r_\delta-R,0) $ and $ (UB_{i,m-1},0) $, respectively. Fig. \ref{joint} (b) provides a detailed illustration of the space $ S_1 $ in Fig. \ref{joint} (a). The probability $ P_{i,m}(o_\delta) $ is calculated by
\begin{equation}
\begin{aligned}
P_{i, m}(o_\delta)
&=\prod_{k=1}^{n_{i,m}}{P_{i,m}(r_\delta \geq r_k)}
\\
&=\prod_{k=1}^{n_{i,m}}{ \frac{\int_{0}^{r_\delta}{(R^2-r^2)dr}}{\int_{0}^{m\cdot R}{(R^2-r^2)dr}} }
\\
&=\left(\frac{{r_\delta}^3}{(m\cdot R)^3}\right)^{n_{i,m}}
\end{aligned}, m>1,
\label{eq4}
\end{equation}
where $ r_\delta=max(r_1,r_2,...,r_k,...,r_{n_m}) $, $ n_{i,m} $ denotes the number of nodes detected by $ a_i $ when $ hop \le m $.

Correspondingly, $ P_{i,m-1}(o_j) $ is calculated by
\begin{equation}
P_{i,m-1}(o_j)=1-\left(1-\frac{CS}{S_{i,m-1}}\right)^\epsilon,
\label{eq5}
\end{equation}
where $ \epsilon $ indicates the node density, and it is calculated by $ \epsilon = \frac{n_{i,m-1} - n_{i,m-2}}{2} $. $ S_{i,m-1} $ denotes the maximum space detected by $ a_i $ when $ hop=m-1 $, it is calculated by
\begin{equation}
S_{i,m-1}=\frac{4}{3}\pi UB_{i,m-1}^3,
\label{eq6}
\end{equation}
and $ CS $ denotes the communal space detected by $ a_i $ and $ o_\delta $, it is composed of $ S_1 $ and $ S_2 $. Based on the analysis of Fig. \ref{joint}, $ CS $ is calculated by
\begin{equation}
\begin{aligned}
CS{\tiny }=&\int_{r_\delta-R}^{r_j}{\pi(R^2-x^2)dx}
\\
&+\int_{r_j}^{E_{i,m-1}(o_\delta)}{\pi(UB_{i,m-1}^2-x^2)dx}
\end{aligned} \ ,
\label{eq7}
\end{equation}
where $ UB_{i,m-1} $ denotes the expected distance of the outermost node detected by $ a_i $ when $ hop=m-1 $. $ UB_{i,m} $ is calculated by
\begin{equation}
UB_{i,m}=\int_{0}^{m\cdot R}{r_\delta{P_{i,m}^{true}(o_\delta)}^\prime d r_\delta}.
\label{eq8}
\end{equation}
Specially, when $ m=1 $, its expected distance is $ \frac{3n_{i,1}}{3n_{i,1}+1}R $. This result indicates that when the number of detected nodes $ n_1 \rightarrow\infty $, the $ UB_{i,1}=R $. It verifies the reliability of our model.

\section{Probability-based Average Distance Estimation}

In this section, based on the upper boundary obtained from the previous section, we propose a probability-based average distance estimation model for 3D sensor localization scenarios. Compared to DEM-DV-Hop\cite{Wang2023}, the PADE model considers the conditional probability during node detection when calculating the average distance. This consideration can enable the PADE model to achieve higher accuracy in distance estimation. Based on the information presented in Fig. \ref{hopm} and the PMDE model, we present the derivation process of the PADE model. Similarly, two conditions need to be satisfied: \textbf{1)} $ {u}_{k}$ is a node within the detection range of $ {a}_{i}$ when $ hop=m $; \textbf{2)} $ {o}_{j}$ is a relay node located in the communal space.

\begin{figure}[t]
	
	\centerline{\includegraphics[scale=0.5]{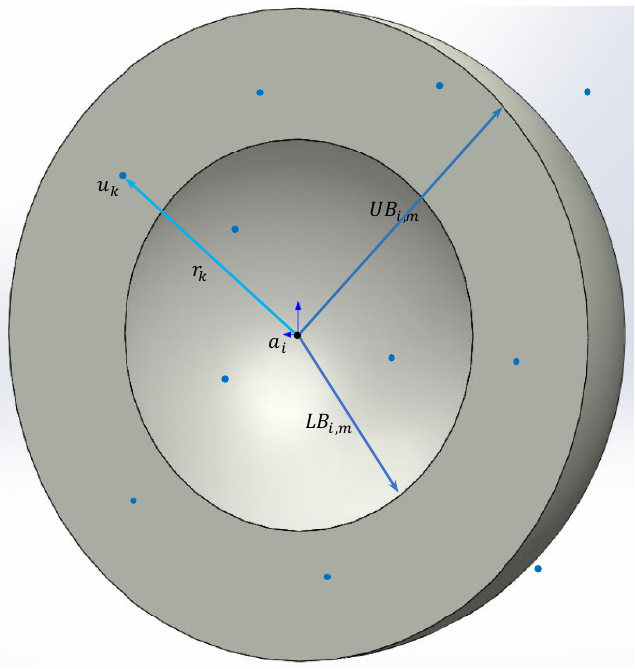}}
	\caption{An example of the distribution space of $ u_k $ when $ m=1 $. $ r_k $: the distance between $ u_k $ and $ {a}_{i}$, $ {UB}_{i,m}$ and $ {LB}_{i,m}$: the upper and lower bounds of $ r_k $.}
	\label{huan}
\end{figure}

Specifically, let $ P_{i, m-1}(o_j) $ represent the probability that there exists at least one $ {o}_{j}$ in the communal space; let $ P_{i,m}(u_k) $ represent the probability that $ {u}_{k}$ is an unknown node within the detection range of $ {a}_{i}$ when $ hop=m $. Let $ P_{i, m}^{true}(u_k) $ represent the probability density that $ {u}_{k}$ can be detected by $ a_i $ when $ hop=m $. and $ P_{i, m}^{true}(u_k) $ is calculated by
\begin{equation}
P_{i, m}^{true}(u_k)=P_{i, m-1}(o_j)P_{i,m}(u_k),
\label{eq11}
\end{equation}
where $ P_{i,m-1}(o_j) $ is calculated by (\ref{eq5}).  

The distribution of $ {u}_{k}$ is shown in Fig. \ref{huan}. In this figure, $ r_k $ denotes the distance between $ u_k $ and $ {a}_{i}$, $ {UB}_{i,m}$ and $ {LB}_{i,m}$ denote the upper and lower bounds of $ r_k $. $ P_{i,m}(u_k) $ can be considered as the probability density function of $ {u}_{k}$ within a circular spatial domain, and it is calculated by
\begin{equation}
P_{i, m}(u_k)=\frac{r_k^3-{LB}_{i,m}^3}{{UB}_{i,m}^3-{LB}_{i,m}^3},
\label{eq12}
\end{equation}
where $ {UB}_{i,m}$ is calculated by (\ref{eq8}), Inspired by \cite{caiSC2020} and \cite{Wang2023}, when $ m>2 $, we define the expected distance ($ E\_dis_{i,m} $) of $ {u}_{k}$ detected by $ {a}_{i}$ at $ hop=m-1 $ as $ {LB}_{i,m}$. The $ E\_dis_{i,m} $ is calculated by
\begin{equation}
E\_dis_{i,m}=\int_{{LB}_{i,m}}^{{UB}_{i,m}}{r_k{P_{i,m}^{true}(u_k)}^\prime d r_k},
\label{eq13}
\end{equation}
where  $ {LB}_{i,m}= E\_dis_{i,m-1}$. When $ m=2 $, the value of $ {LB}_{i,m} $ is equal to $ {UB}_{i,1} $. Specially, when $ m=1 $, the value of $ {LB}_{i,m} $ is equal to 0, the result of $ E\_dis_{i,m} $ is $ \frac{3}{4}UB_{i,1} $.

It should be emphasized that the calculation method described in (\ref{eq13}) differs entirely from the approaches outlined in \cite{caiSC2020} and \cite{Wang2023}. In this work, we specifically calculate nodes that fulfill the specified conditions within the defined search space. In contrast, \cite{caiSC2020} and \cite{Wang2023} calculate all nodes within the defined search space. It suggests that our approach has the potential to theoretically offer more precise accuracy in distance estimation.

\section{Multi-Objective Optimization}

In this section, we construct two distance loss functions ($ \mathcal{L}_1 $ and $ \mathcal{L}_2 $) based on $ Dis{_{i,k}} $ and $ E\_dis_{i,k} $ derived from the aforementioned analysis. These distance loss functions are subsequently integrated into multi-objective genetic algorithms\cite{deb2002fast} to predict the location of $ u_k $. This scheme endeavors to deliver precise location prediction solutions for unknown nodes.

\begin{algorithm}[t]
	
	\caption{Multi-objective 3D localization genetic algorithm.} 
	\label{Alg2}
	
	\hspace*{0.02in} {\bf Input:} 
	The locations of anchor nodes and the ground truth 
	\hspace*{0.5in}locations of unknown nodes
	
	\hspace*{0.02in} {\bf Output:} 
	The predicted locations of unknown nodes
	\begin{algorithmic}[1]
		
		\State Initialize parameters with Table \ref{tab1} and initial population
		
		\State Statistical $ hop_{i,k} $ and $ n_{i,m} $ based on DV-Hop algorithm
		
		\State $ Dis_{i,k} $  $\gets $  (\ref{eq01})
		
		\State $ UB_{i,m} $  $\gets $  (\ref{eq8})
		
		\State $ {E\_dis}_{i,k}$ $\gets $  (\ref{eq13})
		
		\For{$ iter= 1 \to M\_Iter$} 
		
		\State Perform crossover and mutation operations based on \hspace*{0.2in}$ Pc $ and $ Pm $.
		
		\State $ \mathcal{L}_1 $ $\gets $ (\ref{f1})
		
		\State $ \mathcal{L}_2 $ $\gets $ (\ref{f2}).
		
		\State Calculate non-dominated sorting and crowding \hspace*{0.22in}distance based on the values of $ \mathcal{L}_1 $ and $ \mathcal{L}_2 $.
		
		\State Select competitive population individuals based on \hspace*{0.22in}sorting results.

		\EndFor
		\State{\bf end} 
	\end{algorithmic}
\end{algorithm}

Based on the $ Dis{_{i,k}} $ obtained from (\ref{eq01}), we can construct the first distance loss function ($ \mathcal{L}_1 $) as follows
\begin{equation}
\begin{aligned}
\min& \;{\mathcal{L}_1}{({x_k},{y_k},{z_k})} =  
\\
&\sum\limits_{i=1}^{N_{a}} {(\sqrt {{{({x_i} - {x_k})}^2} + {{({y_i} - {y_k})}^2} + {{({z_i} - {z_k})}^2} }  - Dis{_{i,k}})^2}
\end{aligned},
\label{f1}
\end{equation}
where $ ({x_i},{y_i},{z_i}) $ and $ ({x_k},{y_k},{z_k}) $ refer to the locations of $ a_i $ and $ u_k $, respectively. Similarly, Based on the $ E\_dis_{i,k} $ obtained from (\ref{eq13}), we can construct the second distance loss function ($ \mathcal{L}_2 $) as follows
\begin{equation}
\begin{aligned}
\min& \;{\mathcal{L}_2}{({x_k},{y_k},{z_k})} =  
\\
&\sum\limits_{i=1}^{N_{a}} {(\sqrt {{{({x_i} - {x_k})}^2} + {{({y_i} - {y_k})}^2} + {{({z_i} - {z_k})}^2}}  - E\_dis_{i,k})^2}
\end{aligned}.
\label{f2}
\end{equation}

Algorithm \ref{Alg2} provides pseudo-code for a multi-objective 3D localization genetic algorithm. Where $ hop_{i,k} $ denotes the hop count between $ u_k $ and $ a_i $, $ n_{i,m} $ denotes the number of nodes detected by $ a_i $ when $ hop \le m $. It should be emphasized that the key innovation and contribution of this paper is the proposal of the PADE model and the PMDE model. The purpose is to provide a solid theoretical foundation for algorithms and enhance the overall localization performance. Therefore, for multi-objective genetic algorithms, we follow the original design scheme. The details of crossover, mutation, selection, and non-dominated sorting operation involved in Algorithm \ref{Alg2} can be found in \cite{deb2002fast}.

\section{Simulation And Analysis}

\subsection{Experimental Setup}

\begin{figure}[htbp]
	
	\center
	\subfigure[]{\includegraphics[scale=0.43]{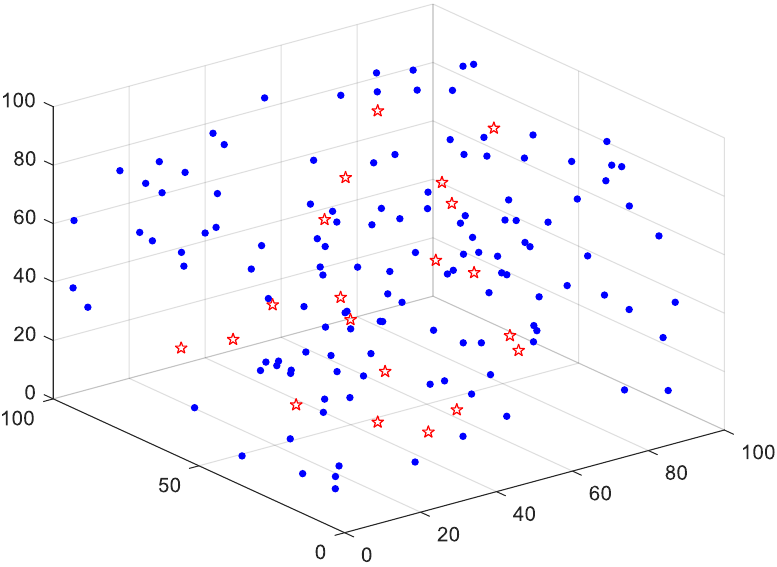}}
	\subfigure[]{\includegraphics[scale=0.43]{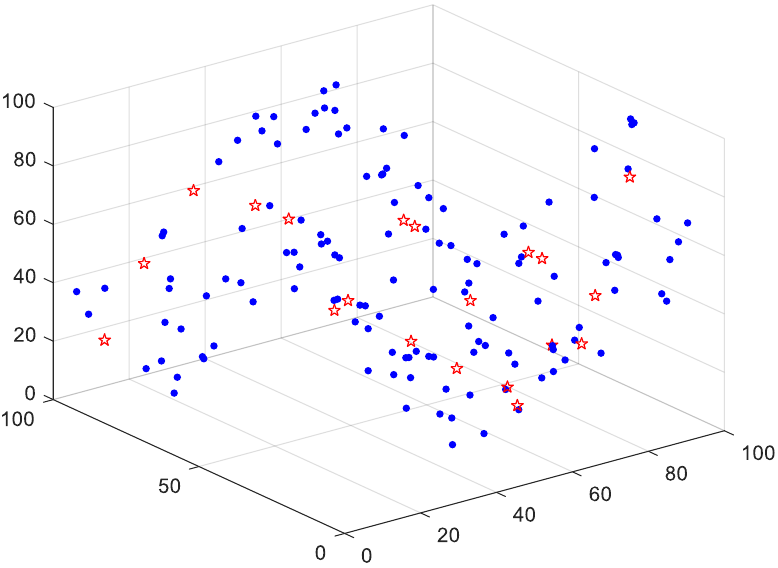}}
	
	\caption{Two different distribution test datasets. (a): random distributed sensor network; (b): multimodal distributed sensor network. \textcolor{red}{\ding{73}}: anchor node, \textcolor{blue}{$ \bullet $}: unknown node.}
	\label{data}
\end{figure}

\textbf{Datasets.} Following\cite{caiS2019}, we chose a randomly distributed network to evaluate localization performance. Moreover, considering the potential deployment of sensors in mountainous environments during real applications, we chose a multimodal distributed sensor network for evaluating the performance. The aim is to assess the generalization ability of the proposed model in adapting to complex deployment environments in practical scenarios. Each sensor network contains the locations of 150 sensor nodes distributed in a 3D space (100m×100m×100m), as shown in Fig. \ref{data}. A subset of the node locations corresponds to the locations of anchor nodes, while another subset corresponds to the ground truth (gt) locations of unknown nodes, corresponding to \textcolor{red}{\ding{73}} and \textcolor{blue}{$ \bullet $} in Fig. \ref{data}, respectively.

\begin{table}[htbp]
	\renewcommand{\arraystretch}{1.15}
	\caption{Simulation parameters. }
	\label{tab1}
	\centering
	\begin{threeparttable}
		\begin{tabular}{p{2.7cm}<{\centering}  p{1.3cm}<{\centering}   p{1.5cm}<{\centering}   p{1.3cm}<{\centering} }
			\hline
			\hline
			\textbf{ Parameters} & \textbf{Value}   &\textbf{ Parameters}		& \textbf{Value}    \\
			
			\hline
			
			$ Pc $      & 0.9 &   $N_{a} $ &   10-35   \\
			
				$ Pm $    & 0.1  & $ N_{u} $     & 115-140  \\
			
			$ Ps $       & 20  &  $ R  $    & 25-40	  \\
			
			Independent repeat test    & 50   &  $ M\_Iter $       & 500 	\\
			
			\hline
			\hline
			
		\end{tabular}
		$ Pc $: crossover probability,   $ Pm $: mutation probability, $ Ps $: Population size, $ N_{u} $: the number of unknown nodes, $ M\_Iter $: maximum iterations, independent repeat test: the number of independent experiments; population size: the number of individuals in the population.
	\end{threeparttable}
\end{table}

\textbf{Parameters.} The detailed experimental parameters of the algorithm are listed in Table \ref{tab1}.

\begin{table*}[t]
	\renewcommand{\arraystretch}{1.15}
	\caption{The $ ALEs $ in the randomly topological networks.}
	\label{randm}
	\centering
	
	\begin{tabular}{ccccccccccccc}
		\hline
		\hline
		\multicolumn{1}{c}{$ {N}_{a} $} & \multicolumn{4}{c}{10} & \multicolumn{4}{c}{15} & \multicolumn{4}{c}{20} \\  
		\cmidrule(lr){1-1}
		\cmidrule(lr){2-5}
		\cmidrule(lr){6-9}
		\cmidrule(lr){10-13}
		
		\multicolumn{1}{c}{ $ R $ (m)} & \multicolumn{1}{c}{25} &  \multicolumn{1}{c}{30} & \multicolumn{1}{c}{35} &  \multicolumn{1}{c}{40} & \multicolumn{1}{c}{25} &  \multicolumn{1}{c}{30} & \multicolumn{1}{c}{35} &  \multicolumn{1}{c}{40} & \multicolumn{1}{c}{25} &  \multicolumn{1}{c}{30} & \multicolumn{1}{c}{35} &  \multicolumn{1}{c}{40} \\
		\hline
		
		DV-Hop 3D\cite{niculescu2003dv09}	&    90.06 	& 62.68 	& 49.15 	& 47.56 	& 74.71 	& 53.59 	& 41.10 	& 35.59 	& 65.52 	& 49.83 	& 36.02 	& 34.60  \\

		OCS-DV-Hop 3D\cite{CuiJ2017} &   \textbf{65.38} 	& 46.74 	& 37.66 	& 35.16 	& \textbf{61.47} 	& 40.29 	& 32.50 	& 32.92 	& 60.38 	& 39.12 	& 31.83 	& 30.06 	\\
		
		NSGAII-DV-Hop 3D\cite{caiS2019} &	94.62 	& 62.79 	& 52.77 	& 47.56 	& 78.98 	& 51.37 	& 38.71 	& 35.25 	& 71.40 	& 44.59 	& 34.14 	& 29.71 	\\
		
		CC-DV-Hop 3D\cite{GuiTVT2020}	&   73.60 	& \textbf{42.11} 	& \textbf{35.84} 	& \textbf{34.54} 	& 69.46 	& 44.31 	& 30.85 	& 29.68 	& 60.58 	& 37.88 	& 30.39 	& 28.67    \\
		
		IAGA-DV-Hop 3D\cite{OUYANG2021}	&   89.12 	& 56.82 	& 47.18 	& 43.62 	& 69.27 	& 45.85 	& 34.12 	& 30.69 	& 64.69 	& 40.12 	& 30.29 	& 25.86 	\\
		
		DEM-DV-Hop 3D\cite{Wang2023} & 102.22	& 63.76 	& 51.10 	& 45.48 	& 78.51 	& 50.01 	& 38.30 	& 32.09 	& 73.68 	& 46.12 	& 33.01 	& 27.32 	\\

		\hline

		\textbf{Proposed method} &  84.11 & 	50.27 & 	41.54 & 	38.49     & 	62.51 &     \textbf{39.88} & 	\textbf{30.50} & 	\textbf{27.33} & 	\textbf{57.88} & 	\textbf{35.08} & 	\textbf{27.85} & 	\textbf{23.62} 		\\
		
		\hline
		\hline
		
		\multicolumn{1}{c}{$ {N}_{a} $} & \multicolumn{4}{c}{25} &  \multicolumn{4}{c}{30} & \multicolumn{4}{c}{35} \\
		\cmidrule(lr){1-1}
		\cmidrule(lr){2-5}
		\cmidrule(lr){6-9}
		\cmidrule(lr){10-13}

		\multicolumn{1}{c}{ $ R $ (m)} & \multicolumn{1}{c}{25} &  \multicolumn{1}{c}{30} & \multicolumn{1}{c}{35} &  \multicolumn{1}{c}{40} & \multicolumn{1}{c}{25} &  \multicolumn{1}{c}{30} & \multicolumn{1}{c}{35} &  \multicolumn{1}{c}{40} & \multicolumn{1}{c}{25} &  \multicolumn{1}{c}{30} & \multicolumn{1}{c}{35} &  \multicolumn{1}{c}{40}\\
		
		\hline
		
		DV-Hop 3D\cite{niculescu2003dv09}	&  74.56 	& 48.98 	& 36.60 	& 34.49 	& 72.73 	& 43.13 	& 31.92 	& 31.72 	& 63.54 	& 41.38 	& 32.93 	& 32.21   \\
		
		OCS-DV-Hop 3D\cite{CuiJ2017} &   58.08 	& 39.97 	& 29.93 	& 28.66 	& 61.93 	& 38.92 	& 29.67 	& 27.47 	& 60.28 	& 39.29 	& 28.96 	& 28.50		\\
		
		NSGAII-DV-Hop 3D\cite{caiS2019} &	62.41 	& 40.47 	& 30.29 	& 26.90 	& 57.62 	& 36.19 	& 27.28 	& 23.67 	& 55.52 	& 35.25 	& 26.37 	& 23.42 	\\
		
		CC-DV-Hop 3D\cite{GuiTVT2020}	&    61.50 	& 35.78 	& 29.65 	& 29.24 	& 56.26 	& 36.13 	& 27.83 	& 26.27 	& 53.71 	& 35.10 	& 28.01 	& 27.26 	\\

		IAGA-DV-Hop 3D\cite{OUYANG2021}	&   57.12 	& 36.78 	& 26.38 	& 23.83 	& 53.33 	& 32.79 	& 23.99 	& 21.12 	& 51.09 	& 31.83 	& 23.04 	& 20.61 	\\
		
		DEM-DV-Hop 3D\cite{Wang2023} & 65.12 	& 41.95 	& 29.47 	& 25.17 	& 60.97 	& 37.73 	& 26.79 	& 22.99 	& 57.22 	& 36.85 	& 26.20 	& 22.26 	\\ 
		
		\hline

		\textbf{Proposed method} & 	\textbf{50.04} & 	\textbf{32.71} & 	\textbf{24.78} & 	\textbf{21.57} & 	\textbf{46.25} & 	\textbf{29.85} & 	\textbf{22.33} & 	\textbf{19.18} & 	\textbf{44.99} & 	\textbf{29.56} & 	\textbf{21.82} & 	\textbf{18.73}  \\ 
		
		\hline
		\hline
	\end{tabular}
\end{table*}

\begin{table*}[t]
	\renewcommand{\arraystretch}{1.15}
	\caption{The $ ALEs $ in the multimodel distributed network.}
	\label{multi}
	\centering
	
	\begin{tabular}{ccccccccccccc}
		\hline
		\hline
		\multicolumn{1}{c}{$ {N}_{a} $} & \multicolumn{4}{c}{10} & \multicolumn{4}{c}{15} & \multicolumn{4}{c}{20} \\  
		\cmidrule(lr){1-1}
		\cmidrule(lr){2-5}
		\cmidrule(lr){6-9}
		\cmidrule(lr){10-13}
		
		\multicolumn{1}{c}{ $ R $ (m)} & \multicolumn{1}{c}{25} &  \multicolumn{1}{c}{30} & \multicolumn{1}{c}{35} &  \multicolumn{1}{c}{40} & \multicolumn{1}{c}{25} &  \multicolumn{1}{c}{30} & \multicolumn{1}{c}{35} &  \multicolumn{1}{c}{40} & \multicolumn{1}{c}{25} &  \multicolumn{1}{c}{30} & \multicolumn{1}{c}{35} &  \multicolumn{1}{c}{40} \\
		\hline
		
		DV-Hop 3D\cite{niculescu2003dv09}	&    98.61 	& 81.91 	& 81.91 	& 65.69 	& 79.74 	& 73.45 	& 57.05 	& 53.29 	& 80.54 	& 65.92 	& 52.02 	& 55.46 	\\

		OCS-DV-Hop 3D\cite{CuiJ2017} &  \textbf{70.78} 	& \textbf{63.10} 	& 68.47 	& 56.61 	& 58.57 	& \textbf{51.61} 	& 43.17 	& 46.70 	& 45.66 	& 42.72 	& 37.00 	& 33.32 	\\
		
		NSGAII-DV-Hop 3D\cite{caiS2019} &	96.45 	& 79.63 	& 75.50 	& 64.23 	& 71.29 	& 62.92 	& 52.35 	& 50.55 	& 58.42 	& 51.34 	& 42.42 	& 41.39 	\\
		
		CC-DV-Hop 3D\cite{GuiTVT2020}	& 76.18 	& 67.89 	& \textbf{58.34} 	& \textbf{49.77} 	& 71.13 	& 68.31 	& 49.17 	& 48.59 	& 67.27 	& 58.49 	& 45.31 	& 47.73 
		\\
		
		IAGA-DV-Hop 3D\cite{OUYANG2021}	&   86.78 	& 77.54 	& 68.64 	& 56.70 	& 62.34 	& 58.04 	& 45.45 	& 45.57 	& 48.95 	& 45.43 	& 36.01 	& 35.07 	\\
		
		DEM-DV-Hop 3D\cite{Wang2023} & 93.03 	& 80.01 	& 72.59 	& 59.06 	& 71.98 	& 64.02 	& 50.32 	& 48.43 	& 57.83 	& 50.76 	& 40.58 	& 38.17 	\\

		\hline

		\textbf{Proposed method} &  76.91 &	71.44 & 	64.58 & 	53.69 & 	\textbf{55.91} & 	52.01 & 	\textbf{42.48} & 	\textbf{41.29} & 	\textbf{42.24} & 	\textbf{38.65} & 	\textbf{32.86} & 	\textbf{32.03} 	\\
		
		\hline
		\hline
		
		\multicolumn{1}{c}{$ {N}_{a} $} & \multicolumn{4}{c}{25} &  \multicolumn{4}{c}{30} & \multicolumn{4}{c}{35} \\
		\cmidrule(lr){1-1}
		\cmidrule(lr){2-5}
		\cmidrule(lr){6-9}
		\cmidrule(lr){10-13}

		\multicolumn{1}{c}{ $ R $ (m)} & \multicolumn{1}{c}{25} &  \multicolumn{1}{c}{30} & \multicolumn{1}{c}{35} &  \multicolumn{1}{c}{40} & \multicolumn{1}{c}{25} &  \multicolumn{1}{c}{30} & \multicolumn{1}{c}{35} &  \multicolumn{1}{c}{40} & \multicolumn{1}{c}{25} &  \multicolumn{1}{c}{30} & \multicolumn{1}{c}{35} &  \multicolumn{1}{c}{40}\\
		
		\hline
		
		DV-Hop 3D\cite{niculescu2003dv09}	&  75.53 	& 64.26 	& 49.22 	& 52.10 	& 72.41 	& 61.88 	& 51.56 	& 52.24 	& 66.65 	& 55.12 	& 49.45 	& 46.05 	\\
		
		OCS-DV-Hop 3D\cite{CuiJ2017} &   45.48 	& 43.85 	& 37.69 	& 35.55 	& 47.69 	& 39.44 	& 36.20 	& 33.76 	& 50.34 	& 41.74 	& 37.87 	& 34.37 
		\\
		
		NSGAII-DV-Hop 3D\cite{caiS2019} &	54.40 	& 47.99 	& 40.79 	& 40.07 	& 51.17 	& 45.61 	& 38.86 	& 38.03 	& 50.63 	& 45.21 	& 38.92 	& 38.08 	\\
		
		CC-DV-Hop 3D\cite{GuiTVT2020}	& 59.67 	& 51.77 	& 39.88 	& 41.80 	& 53.73 	& 49.13 	& 42.23 	& 38.42 	& 51.50 	& 44.99 	& 39.74 	& 37.84 
		\\

		IAGA-DV-Hop 3D\cite{OUYANG2021}	&   45.41 	& 41.23 	& 34.40 	& 33.33 	& 45.03 	& 39.66 	& 33.63 	& 32.00 	& 43.89 	& 38.82 	& 33.36 	& 31.43 
		\\
		
		DEM-DV-Hop 3D\cite{Wang2023} & 51.82 	& 46.43 	& 37.98 	& 36.02 	& 49.48 	& 44.61 	& 36.80 	& 34.11 	& 48.41 	& 43.95 	& 36.70 	& 33.90 	\\ 
		
		\hline

		\textbf{Proposed method} & 	\textbf{39.09} & 	\textbf{36.13} & 	\textbf{32.38} & 	\textbf{31.06} & 	\textbf{37.08} & 	\textbf{34.16} & 	\textbf{31.63} & 	\textbf{29.42} & 	\textbf{36.77} & 	\textbf{33.40} & 	\textbf{31.72} & 	\textbf{28.99} \\ 
		
		\hline
		\hline
	\end{tabular}
\end{table*}

\begin{table}[t]
	\renewcommand{\arraystretch}{1.15}
	\caption{Comparison of comprehensive localization performance. }
	\label{ALA}
	\centering
	\begin{threeparttable}
		\begin{tabular}{cccccccccc}
			
			\hline
			\hline
			\multicolumn{1}{c}{Node distribution type} & \multicolumn{2}{c}{Random} & \multicolumn{2}{c}{Multimodal}  \\ 
			
			\cmidrule(lr){1-1}
			\cmidrule(lr){2-3}
			\cmidrule(lr){4-5}

			Algorithms & \textbf{$ ALA $ }  & \textbf{$ APG $ }   & \textbf{$ ALA $ }  & \textbf{$ APG $ }    \\
			
			\hline
			
			DV-Hop 3D\cite{niculescu2003dv09}  & 50.64      &  12.66  $ \uparrow $   &  35.75  &  22.34 $ \uparrow $  \\
			
			OCS-DV-Hop 3D\cite{CuiJ2017}  & 58.95      &  4.35  $ \uparrow $   &  54.10  &  3.99 $ \uparrow $  \\
			
			NSGAII-DV-Hop 3D\cite{caiS2019}    & 54.70  &  8.60  $ \uparrow $&  46.82   &  11.27 $ \uparrow $  \\

			CC-DV-Hop 3D\cite{GuiTVT2020}   &  59.81      &  3.49  $ \uparrow $&  47.55   &  10.54 $ \uparrow $  \\
			
			IAGA-DV-Hop 3D\cite{OUYANG2021}    & 59.19      &  4.11  $ \uparrow $&  53.39   &   4.70 $ \uparrow $  \\
			
			DEM-DV-Hop 3D\cite{Wang2023}    & 54.40    &  8.9 $ \uparrow $ &  48.87   &  9.22 $ \uparrow $   \\
			
			\hline
			
			\textbf{Proposed method}    & \textbf{63.30}  &  0.00 &  \textbf{58.09}  &  0.00   \\

			\hline
			\hline
		\end{tabular}
		
	\end{threeparttable}
\end{table}

\textbf{Performance Metric.} In this work, we provide multiple metrics to comprehensively evaluate the effectiveness of the proposed method. Firstly, we employ the normalized average localization errors ($ ALEs $) as a metric to assess the algorithm's localization errors in various testing scenarios. This metric aims to evaluate the generalization capability of the proposed algorithm. The $ ALEs $ are calculated by
\begin{equation}
ALEs = \frac{100\%}{{N_{u} R}}\sum\limits_{k = 1}^{N_{u}} {\sqrt {{{(x_k^{gt} - {x_k})}^2} + {{(y_k^{gt} - {y_k})}^2} + {{(z_k^{gt} - {z_k})}^2}} },
\label{eq18}
\end{equation}
where $ (x_k,y_k,z_k) $ and $ (x_k^{gt},y_k^{gt},z_k^{gt}) $ represent the prediction coordinates and ground truth coordinates of an $ u_k $ respectively. Next, the average performance gain ($ APG $) is used to evaluate the comprehensive localization gain of the proposed algorithm compared to other algorithms. This metric focuses on evaluating the effectiveness of the proposed algorithm. The $ APG $ is calculated by
\begin{equation}
APG= {\rm mean}\sum{({ ALEs}_{other}-{ALEs}_{our})}, 
\label{eq15}
\end{equation}
where $ {ALEs}_{other} $ represents the $ ALEs $ of other algorithms used for comparison in this paper. Lastly, we employ the $ 95\% $ confidence interval of $ ALEs $ ($ ALEs $ follow a Gaussian distribution) as a mean to visualize the distribution of errors. This metric aims to evaluate the stability of the proposed algorithm. The confidence interval is calculated by
\begin{equation}
[\bar{X}-\frac{S}{\sqrt n}t_{{\frac{\alpha}{2}}}(n-1), \bar{X}+\frac{S}{\sqrt n}t_{{\frac{\alpha}{2}}}(n-1)], 
\label{eqint}
\end{equation}
where $ \bar{X} $, $ S^2 $, $ n $, and $ \alpha $ denote the mean, variance, and samples size, and significance level, respectively. In this work, $ \bar{X}={\frac{1}{n}}\sum_{i}^{n}{ALE}_i $, $ S^2=\frac{\sum_{i}^{n}{({ALE}_i-\bar{X})}^2}{n-1} $, $ n=50 $, and $\alpha =0.05 $.

\subsection{Comparison with State-of-the-art Results}

To objectively showcase the effectiveness and stability of the proposed method, we compare several baseline algorithms. Specifically, DV-Hop 3D\cite{niculescu2003dv09} is a classic range-free localization algorithm. DEM-DV-Hop 3D\cite{Wang2023} is an algorithm that uses distance estimation strategy and represents the state-of-the-art in 2D positioning scenes. OCS-DV-Hop 3D\cite{CuiJ2017} and IAGA-DV-Hop 3D\cite{OUYANG2021} are algorithms based on enhanced cuckoo search and genetic algorithm for position prediction, respectively. CC-DV-Hop 3D\cite{GuiTVT2020} is a classic method for transforming the localization problem into constrained optimization problem.  NSGAII-DV-Hop 3D\cite{caiS2019} is a classic multi-objective localization scheme in 3D positioning scenes.

Table \ref{randm} and Table \ref{multi} provide performance comparisons with current excellent works under random distributed sensor network and a multimodal distributed sensor network, respectively. In general, the $ ALEs $ of the proposed method demonstrate a consistent trend across various test datasets, whereby the $ ALEs $ decrease as the number of anchor nodes increases. Meanwhile, the $ ALEs $ decrease with the increase of node communication radius. In addition, the results demonstrate that the $ ALEs $ of the proposed method outperforms DV-Hop 3D\cite{niculescu2003dv09}, DEM-DV-Hop 3D\cite{Wang2023}, IAGA-DV-Hop 3D\cite{OUYANG2021}, and NSGAII-DV-Hop 3D\cite{caiS2019} across all test conditions. When $ N_a=10 $, the $ ALEs $ of the proposed method is more outstanding than OCS-DV-Hop 3D\cite{CuiJ2017} and CC-DV-Hop 3D\cite{GuiTVT2020}; when $ N_a>10 $, the $ ALEs $ of the proposed method is superior to these two algorithms across almost all conditions. In summary, based on the results presented in Tables \ref{randm} and \ref{multi}, it is evident that the proposed method has achieved significant gains compared to the comparative algorithm.

Table \ref{ALA} displays the average localization accuracy ($ ALA $, $ ALA =1- mean(ALEs) $) of the algorithm and the $ APG $ achieved by the proposed method in comparison to other algorithms across two different test data. The results indicate that the proposed method achieves an $ ALA $ of $ 63.30\% $ and $ 58.09\% $ in random and multimodal distributed sensor networks, respectively. Compared to the state-of-the-art algorithms, the proposed method achieves $ APG $ ranging from $ 3.49\% $ to $ 12.66\% $ in the random distributed sensor network and from $ 3.99\% $ to $ 22.34\% $ in the multimodal distributed sensor network. It is noteworthy that in 2D localization scenarios, \cite{Wang2023} achieves the best localization performance in comparison algorithms. However, in 3D localization scenarios, its comprehensive localization performance is inferior to algorithms such as \cite{CuiJ2017}, \cite{GuiTVT2020}, and \cite{OUYANG2021}. This phenomenon highlights the infeasibility of transferring probability-based distance estimation models from 2D environments to their 3D counterparts without appropriate modifications. It demonstrates the significance of this work.

\begin{figure}[t]
	
	\center
	\subfigure[]{\includegraphics[scale=0.45]{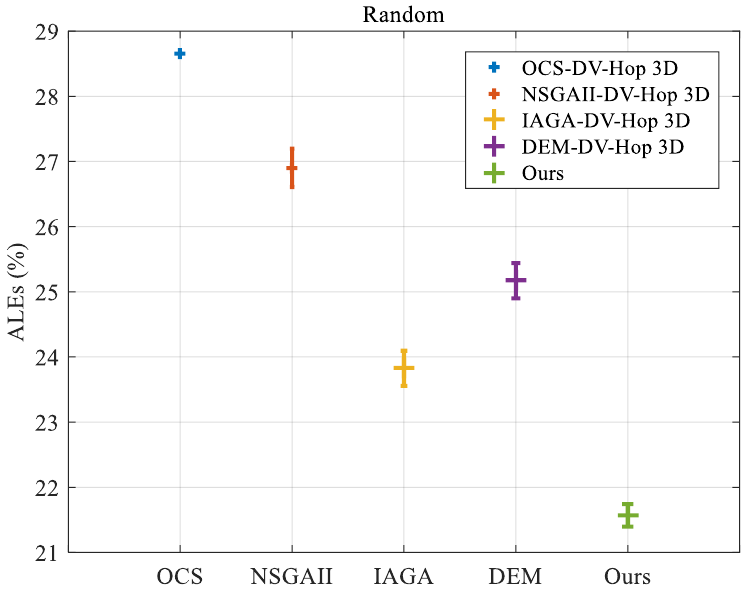}}
	\subfigure[]{\includegraphics[scale=0.45]{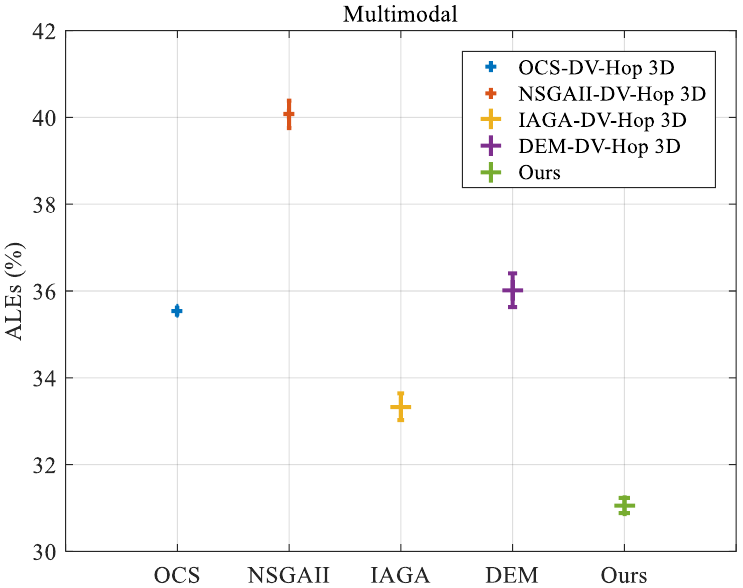}}
	
	\caption{The $ 95\% $ confidence interval for $ ALEs $ samples. (a): random distributed sensor networks; (b): Multimodal distributed sensor networks. $ N_a=25 $ and $ R=40 $.}
	\label{std}
\end{figure}

Fig. \ref{std} provides a visualization of the $ 95\% $ confidence interval (CI) for the $ ALEs $ samples, considering the parameters $ N_a=25 $ and $ R=40 $. In this study, a consistent confidence level and sample size were maintained throughout the testing process. As a result, the width of the CI accurately reflects the dispersion level of samples. The results demonstrate that the CI width of the proposed method is slightly wider than OCS-DV-Hop 3D\cite{CuiJ2017}, but the localization performance is significantly better than OCS-DV-Hop 3D\cite{CuiJ2017}. In addition, the CI width of the proposed method is significantly better than other comparative algorithms. The results of CI indicate that the proposed method demonstrates favorable stability and reliability. The predicted location of each node obtained by DV-Hop 3D\cite{niculescu2003dv09} and CC-DV-Hop 3D\cite{GuiTVT2020} is unique and thus not considered in the comparison of the CI.

\section{Conclusions}

In 3D localization studies, most algorithms focus on enhancing position estimation algorithms, lacking theoretical derivation of the detection distance of an anchor node at the varying hops, which engenders a localization performance bottleneck. To address this issue, we propose a PADE model for 3D DV-Hop localization in WSNs. The aim is to mathematically derive the average distances of nodes detected by an anchor node at different hops. First, we develop a probability-based maximum distance estimation (PMDE) model to calculate the upper bound of the distance detected by an anchor node. Then, we present the PADE model  relies on the upper bound obtained of the distance by the PMDE model. Finally, the obtained average distance is used to construct a distance loss function, and it is embedded with the traditional distance loss function into a multi-objective genetic algorithm to predict the locations of unknown nodes. The experimental results demonstrate the proposed method achieves the state-of-the-art performance in both random and multimodal distributed sensor networks.

In future work, we will investigate obstacle detours in 3D positioning scenarios and apply the derived method to diverse complex deployment environments.

\ifCLASSOPTIONcaptionsoff
\newpage
\fi

\bibliographystyle{IEEEtran}
\bibliography{refs}

\end{document}